\documentclass[a4paper]{article}
\usepackage{RR}
\usepackage{hyperref}

\usepackage{amsmath}
\usepackage{amssymb}
\usepackage{subfigure}
\usepackage{ifthen}
\usepackage[usenames]{color}
\usepackage{framed}
\usepackage{tabularx}

\newtheorem{definition}{Definition}

\newtheorem{theorem}{\mbox{Theorem}}

\newtheorem{lemma}{\mbox{Lemma}}
\newenvironment{proof}{%
\vspace*{0.1cm}\noindent{ {Proof} : } \\ \it}%
{}

\RRdate{Novembre 2008}

\newboolean{Notes}

\newcommand{\Notes}[1]
{
  \ifthenelse{\boolean{Notes}}
  {\definecolor{shadecolor}{gray}{0.80} \begin{shaded} {\small #1} \end{shaded}}{}
}

\setboolean{Notes}{false}
\newcommand{\cqfd}{\hfill $\Box$}

\RRauthor{
Beno\^it Delahaye, Universit\'e de Rennes 1 / IRISA

  \and
Beno\^it Caillaud,  INRIA / IRISA 

}
\authorhead{B. Delahaye \& B. Caillaud}
\RRtitle{Un mod\`ele de raisonnement probabiliste bas\'e sur les contrats Assume/Guarantee}
\RRetitle{A Model for Probabilistic Reasoning on Assume/Guarantee Contracts}

\RRabstract{In this paper, we present a probabilistic adaptation of an Assume/Guarantee contract formalism. For
the sake of generality, we assume that the extended state machines used in the contracts and implementations define
sets of runs on a given set of variables, that compose by intersection over the common variables. In
order to enable probabilistic reasoning, we consider that the contracts dictate how certain
input variables will behave, being either non-deterministic, or probabilistic; the introduction of
probabilistic variables leading us to tune the notions of implementation, refinement and composition. As
shown in the report, this probabilistic adaptation of the Assume/Guarantee contract theory preserves
compositionality and therefore allows modular reliability analysis, either with a top-down or a
bottom-up approach.
}

\RRkeyword{
Assume/Guarantee Reasoning, Contracts, Probabilistic reasoning, Reliability analysis.
}

\RRresume{
Ce document pr\'esente une adaptation probabiliste d'un formalisme de contrats {\it Assume/Guarantee}. On
supposera, dans le but d'\^etre le plus g\'en\'eral possible, que les machines à états étendues utilis\'ees pour les contrats
et impl\'ementations d\'efinissent des ensembles d'histoires sur un ensemble de variables donné, et qu'elles se
composent par intersection sur les variables communes. Pour permettre un raisonnement probabiliste,
on consid\`ere que les contrats pr\'ecisent le comportement des variables d'entr\'ees, non-d\'eterministes ou
probabilistes. Le fait de consid\'erer des entr\'ees probabilistes n\'ecessite une adaptation des notions
d'impl\'ementation, composition et raffinement. Ce rapport montre que cette adaptation probabiliste de la th\'eorie
des contrats {\it Assume/Guarantee} pr\'eserve la compositionalit\'e, et permet de ce fait une analyse de
fiabilit\'e modulaire, que ce soit par une approche ascendante ou descendante.

}

 \RRmotcle{
 Raisonnement Assume/Guarantee, Contrats, Raisonnement probabiliste, Analyse de fiabilit\'e.
 }

\RRprojet{S4}  
\RRtheme{\THCom \THSym} 
 \RCRennes 

\begin{document}
\RRNo{6719}

\makeRR   

\section{Introduction}

Several industrial sectors involving complex embedded systems
have recently experienced deep changes in their organization,
aerospace and automotive being the most prominent examples.
In the past, they were organized around vertically integrated
  companies, supporting in-house design activities.
 These sectors have now evolved into more specialized,
  horizontally structured companies: equipment suppliers and OEMs.
OEMs perform system design and integration by importing/reusing
  entire subsystems provided by equipment suppliers.
As a consequence, part of the design load has been moved from OEMs to
  suppliers. An inconvenient of this change is the increased occurrence of late error discovery, system level
  design errors uncovered at integration time.
This is particularly true for system reliability, for state of
  the art reliability analysis techniques are not modular~\cite{HOY,SA96}.

\Notes{ Références?}

 A corrective action, taken in the last decade is that the OEMs now focus on the
  part of the system design at the core of their business, and as far
  as possible, rely on industry-wide standard platforms. This has an impact on design methods and modeling formalisms: Virtual
  prototyping, design space exploration are required early in the
  design cycle. Component based design has emerged as the most
  promising technique to address the challenges resulting from this
  new organization of the industry.

 However, little has been done regarding the capture of
  reliability requirements, their formalization in behavioural models
  and the verification techniques capable of analyzing in a modular
  way the reliability aspects of a system, at an early stage of design.
 The paper contributes to solve these issues: The semantics
  foundations presented in this paper consists in a mathematical formalism
  designed to support a component based design methodology and to
  offer modular and scalable reliability analysis techniques.
 At its basis, the mathematical formalism is a language theoretic
  abstraction of systems behaviour. This basic formalism can be
  instantiated to cover several aspects, including functional,
  timeliness, hybrid and reliability~\cite{BEN2007}. This report presents the
  reliability aspect.

 The central concept of the formalism is the notion of contract,
  built on top of a basic behavioural formalism. Contracts allow to
  distinguish hypotheses on a component from hypotheses made on its
  environment. Contracts are central to component based design
  methodologies.

 This paper focuses on developing a compositional theory of
  probabilistic contracts, capable of capturing reliability aspects of
  components and systems. The key contributions are the definition of probabilistic satisfaction,
  composition and refinement relations that ensure that they will be compositional.

 The paper is organized as follows: In the first section we present the Assume/Guarantee
 formalism upon which we built our probabilistic theory. In the second section we formally define
 the probabilistic Assume/Guarantee theory we developed and we prove that it is compositional. In a
 third section we compare our work to classical related formalisms like pCTL and pCTL$^*$~\cite{ASSB95,HJ89}, or developed more recently, such as  Dynamic Fault
 Trees~\cite{BoudaliCS07} and Arcade~\cite{BoudaliICECCS08}.

\Notes{Ref Stoelinga 

En mettre un peu plus sur les related work

}


\section{Classical Assume/Guarantee Reasoning}

The model we have built is based upon the notion of components and Assume/Guarantee reasoning. In
this section, we will define the background upon which the report is based. In a first subsection we will give the
definitions we used for contracts and implementations. Then we will present the basic
operations already existing on contracts and the main theorems we want to preserve in our
probabilistic adaptation.

\Notes{
A etoffer?
}

\subsection{Contracts and Implementations}

In order to define contracts and implementations, we need to consider the abstract notion of
``assertion''.

\begin{definition}
An assertion $E = S::\sigma$ possesses a set of {\it ports} and {\it variables} (its signature, $\sigma$) through which it interacts with
its environment. $S$ is identified  with the set of runs it defines or accepts, each run assigning a history for
each variable and port. If necessary, the inverse projection of $E =
S::\sigma$ on $\sigma' \supseteq \sigma$ will be denoted by $E\uparrow^{\sigma'}$.
\end{definition}%

We assume that there exists a complementation operator for an assertion $E$, relative to its 
signature $\sigma$. It is denoted by $\lnot
E$. Assertions compose by intersection over the common sets of ports and variables
(assuming the appropriate inverse projections have been performed to
equalize the involved signatures). We will denote products either by
$E_1 \times E_2$ or $E_1 \cap E_2$ equivalently.
$$ E_1 \times E_2 =
E_1 \cap E_2 = S_1\uparrow^\sigma \cap S_2\uparrow^\sigma :: \sigma, \
\mbox{with} \ \sigma = \sigma1 \sup \sigma 2$$

With these notations and definitions, we will be able to define implementations and contracts.

\begin{definition}
An implementation is an assertion, i.e. a set of runs with a given signature.
\end{definition}%

We will use the symbol $M = S_M::\sigma_M$ to refer to implementations. They are ordered by inclusion over the runs
they contain (one more time assuming that the appropriate inverse projections have been
performed). We will say
that an implementation $M$ refines an implementation $M'$ with respect to the signature $\sigma$, written $M
\preceq^\sigma M'$, if and only if
$S_M \subseteq^{\sigma} S_{M'}$, i.e. $S_M\uparrow^\sigma \subseteq S_{M'}\uparrow^\sigma$.

Composition preserves implementation refinement.

\begin{definition}
A contract $C::\sigma$ is a pair of assertions
$(A::\sigma_A,G::\sigma_G)$ with $\sigma \supseteq \sigma_A \cup \sigma_G$.
\begin{itemize}
\item $A$ is the {\bf assumption};
\item $G$ is the {\bf guarantee}, i.e. the promised behavior, under the
  hypothesis that $A$ holds.
\end{itemize}
\end{definition}%

Note that for a contract $C::\sigma = (A::\sigma_A,G::\sigma_G)$, we can
consider the equivalent contract $C'::\sigma =
(A\uparrow^\sigma::\sigma, G\uparrow^\sigma::\sigma)$. Whenever
convenient, we will thus suppose that both assertions of a contract $C$ have
the same signature and we will denote respectively the assumption, the
guarantee and the common signature of $C$ by $A_C, G_C$ and $\sigma_C$.

The following definition of satisfaction will precise the interpretation we make of a contract.

\begin{definition}
An implementation $M$ satisfies a contract $C::\sigma_C=(A_C,G_C)$ (written $M \models C$) if and only if
$$M \cap A_C \subseteq^{\sigma_C} G_C$$
\end{definition}%

Satisfaction can be checked using equivalent formulas:

$$ M \models C \iff M\subseteq G_C \cup \lnot A_C \iff M \cap (A_C \cap \lnot G_C) = \emptyset $$

From these equivalent definitions, we can show that there exists a unique maximal implementation
$M_C$ satisfying a contract $C$:

$$M_C = (G_C \cup \lnot A_C)::\sigma_C$$

This maximal implementation is to be interpreted as the implication $A \Rightarrow C$. We can prove
that an implementation $M$ satisfies contract $C=(A_C, G_C)$ if and only if it satisfies the equivalent
contract $(A_C, M_C)$, and if and only if $M \preceq^{\sigma_C} M_C$. We will say that a contract
$C::\sigma = (A::\sigma_A,G::\sigma_G)$ is in {\bf canonical form} when $G = M_C$, or equivalently when $\lnot A \subseteq G$
or $G \subseteq \lnot A$, and $\sigma = \sigma_A = \sigma_G$. As the canonical form of a contract is unique and the satisfaction of the
contract equivalent to the satisfaction of its canonical form, we will consider only contracts in
canonical form in the rest of this document.

\subsection{Operations on Contracts}

The notion of composition of contracts formalizes how contracts attached to different components
of a system should be combined in order to represent one single component. If $C_1::\sigma_1 = (A_1, G_1)$ and
$C_2::\sigma_2 = (A_2,G_2)$ are two contracts defined as in the previous section, their composition should
respect some rules. First, their promises should be composed, as we want to guarantee that both
$G_1$ and $G_2$ must be respected. Remember that composition is the intersection of the two
assertions, after computing the appropriate inverse projections in order to equalize the sets of variables
and ports.

Regarding the assumptions, we also want to assume that both $A_1$ and $A_2$ are respected, but we
must consider the case when the second contract guarantees that part of the assumption of the first
one is respected and vice-versa. A run satisfying the assumption of the composition should
consequently either satisfy both $A_1$ and $A_2$ or be made unacceptable by the composition of the
guarantees. Thus the following definition:

\begin{definition}
Let $C_1::\sigma_1 = (A_1, G_1)$ and $C_2::\sigma2 = (A_2, G_2)$ be contracts, we define $C_1 \parallel C_2$ to be the
contract $C::\sigma = (A,G)$ such that:
\begin{itemize}
\item $\sigma = \sigma_1 \cup \sigma_2$;
\item $A = (A_1\uparrow^\sigma \cap A_2\uparrow^\sigma) \cup \lnot (G_1\uparrow^\sigma \cap G_2\uparrow^\sigma)$;
\item $G = G_1\uparrow^\sigma \cap G_2\uparrow^\sigma$.
\end{itemize}
\end{definition}%

Remark that the so defined contract is in canonical form.

With the above definition, we can prove that the composition preserves the implementation relation.

\begin{lemma}%
If $M_1 \models C_1$ and $M_2 \models C_2$ then $M_1 \times M_2 \models C_1 \parallel C_2$.
\end{lemma}%

\begin{proof}%
  As the two contracts are supposed in canonical form, we have $M_i \subseteq^{\sigma_i} G_i$. Thus
  $M_1 \times M_2 \subseteq^{\sigma_1 \cup \sigma_2} G_1 \cap G_2$, and $M_1\times M_2 \models C_1
  \cap C_2$.

\cqfd
\end{proof}%

Next, we will need to build a
refinement relation. Intuitively, this relation must be compatible with the
composition operation and the implementation relation. We will thus say that a contract $C$ refines
another contract $C'$ if it assumes less and guarantees more:

\begin{definition}
The contract $C::\sigma = (A,G)$ refines the contract $C'::\sigma' = (A', G')$, written $C \preceq C'$, if and only
if $\sigma \subseteq \sigma'$, $A \supseteq^{\sigma'} A'$ and $G \subseteq^{\sigma'} G'$.
\end{definition}%

We can now prove the following properties (the proof is quite straightforward and
left to the reader):

\begin{lemma}%
If $M$ is an implementation, $C_1$, $C_2$, $C_3$ and $C_4$ four contracts,

\begin{enumerate}
\item If $M \models C_1$ and $C_1 \preceq C_2$, then $M \models C_2$. 
\item If $C_1 \preceq C_2$ and $C_3 \preceq C_4$ then $C_1 \parallel C_3 \preceq C_2 \parallel C_4$.
\end{enumerate}

\end{lemma}%


\section{Extension to Probabilistic Approach}





In this section we will adapt the definitions presented above in order to be able to
express probabilistic properties, like reliability, while preserving compositionality.

What we want to express is the affirmation ``This particular implementation
$M$ satisfies the given contract $C$ with level $\alpha$'', meaning that given the information of
the contract $C$, we can prove that the (probabilistic) measure of the runs of $M$ that do not satisfy the
contract $C$ (i.e. that are within the assumptions but outside of the guarantees) is not above $1 -
\alpha$. More precisely, we still want to consider non-probabilistic assertions, but we want to be
able to express that the environment may induce randomness in our assertions. We must therefore
precise which of the variables/ports associated to the contract are controlled (internal variables for
instance) or uncontrolled (controlled by the environment). We can then
choose a subset of the uncontrolled ports to be subject to probability distributions. There will
then remain a subset of the uncontrolled ports that we will consider non-deterministic. As a
consequence, the signature of each assertion will be divided into two disjoint sets of ports, controlled or uncontrolled $\sigma = \mathbf{u}
\uplus \mathbf{c}$ (note that for a contract, there will only be one such signature).

\subsection{Probabilistic Contracts, Implementations and Satisfaction}

\begin{definition}
A probabilistic contract is a tuple $\mathcal{C} = (C,\mathbf{p},\mathbb{P})$ with
\begin{itemize}
\item $C = (\mathbf{u},\mathbf{c},A,G)$ a non-probabilistic contract;
\item $\mathbf{p} \subseteq \mathbf{u}$ a set of uncontrolled ports;
\item $\mathbb{P}$ a probability distribution over the set of all histories of $\mathbf{p}$.
\end{itemize}
\end{definition}%

Note that the probability distribution is attached to the contract itself and not to the
implementation. We can therefore give our contract to a supplier saying ``Knowing that the histories
of the ports of $\mathbf{p}$ will follow this distribution, can you build an implementation ensuring
that $90 \% $ of the runs will satisfy the contract?''.
Let's now formally define this probabilistic satisfaction relation:

\begin{definition}
An implementation $M$ satisfies probabilistic contract $\mathcal{C}=(C,\mathbf{p},\mathbb{P})$ with
level $\beta$ (written $M \models_\beta \mathcal{C}$) iff $\mathbf{u}_M \subseteq \mathbf{u}_C, \mathbf{c}_M = \mathbf{c}_C$ and
$$\mathcal{M}(M \subseteq G_\mathcal{C}) \ge \beta$$

N.B.: We still consider that $C$ is in canonical form.
\end{definition}%

The predicate $M \subseteq G_C$ is a reference to the set of all histories of the ports in
$\mathbf{p}$ that ensure the induced behaviors of the implementation $M$ (i.e. the inverse
projection on the set of runs of $M$) is included in the guaranteed behaviors, whatever other
non-deterministic choices have been made. Formally:
If $\omega$ is one possible history of the ports in $\mathbf{p}$, we call $\Omega$ the set of all
such histories,

$$\mathcal{M}(M \subseteq G_\mathcal{C}) = \mathbb{P}(\{\omega \in \Omega \mid \{w\} \cap M
\subseteq^{\sigma_\mathcal{C}} G_\mathcal{C} \})$$

This means that we measure the set of histories of $\mathbf{p}$ that ensure that the runs of $M$
will be included in the guaranteed behavior, whatever non-deterministic choices are.
Finally, $M \models_\beta \mathcal{C}$ means that the probability that $M \models C$ w.r.t. the distribution
of histories on the probabilistic ports is higher than $\beta$, whatever non-deterministic choices are.

\Notes{Après réflexion, $\mathbb{P}$ est une probabilité sur les histoires des variables de
  $\mathbf{p}$, donc on a pas besoin de redéfinir une mesure sur cet espace... ?}

\subsection{Probabilistic Composition}

The probabilistic point of view makes it more complicated to compose contracts. As the
distributions on the probabilistic ports are linked to the contracts, what we absolutely do not want
is to compose two contracts whose probabilistic ports overlap. Moreover we also want to avoid the
case where a probabilistic port for the first contract is controlled by the second one (i.e. the
first considers an input port as probabilistic, but this port is an output of the second). In order
to keep a quite simple definition, and because we think this is not too major a restriction, we will
only define the composition of two contracts when they have compatible sets of
controlled/uncontrolled ports (i.e. $\mathbf{c}_1 \cap \mathbf{c}_2 = \emptyset$). Thus the following definition:

\begin{definition}
If  $\mathcal{C}_1$ and $\mathcal{C}_2$ are 2 probabilistic contracts, their parallel composition
$\mathcal{C}_1 \parallel \mathcal{C}_2$ is defined if and only if
\begin{enumerate}
\item $C_1 \parallel C_2$ is defined (i.e. $\mathbf{c}_1\cap \mathbf{c}_2 = \emptyset$);
\item $\mathbf{p}_1$ and $\mathbf{p}_2$ are {\bf disjoint} sets of {\bf uncontrolled} ports in $C_1
  \parallel C_2$ (i.e.  ports that are neither
  controlled by $C_1$ or $C_2$).
\end{enumerate}
Then we have
$$ \mathcal{C}_1 \parallel \mathcal{C}_2 = (C,\mathbf{p},\mathbb{P}) \ \mbox{with} \ 
\left \{ \begin{array}{c@{\ = \ }c}
C & C_1 \parallel C_2 \\
\mathbf{p} & \mathbf{p}_1 \uplus \mathbf{p}_2 \\
\mathbb{P} & \mathbb{P}_1 \times \mathbb{P}_2 \end{array} \right .$$

\end{definition}%

\begin{figure}[h]
  \begin{center}
    \input{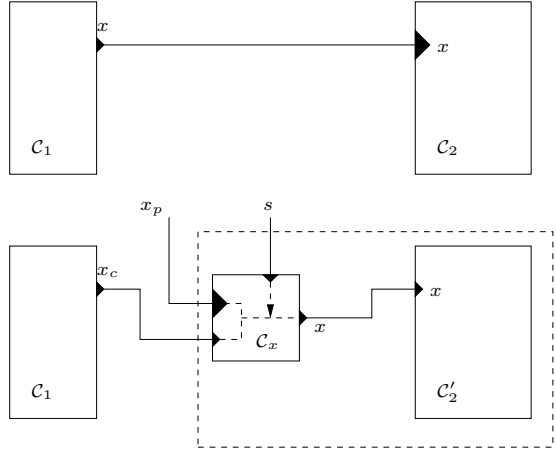}
  \end{center}
  \caption{Illustrating the wrapper mechanism}
  \label{wrapper}
\end{figure}

The above definition makes it impossible to compose two contracts whose probabilistic and controlled
ports overlap. This could be seen as a major restriction but there is a way to make such contracts
compatible. Consider two probabilistic contracts $\mathcal{C}_1$ and $\mathcal{C}_2$, and suppose that
the port $x$ is controlled by $\mathcal{C}_1$, but considered as probabilistic by
$\mathcal{C}_2$. If we want to compose $\mathcal{C}_1$ and $\mathcal{C}_2$, we have to make the port
$x$ non-probabilistic in $\mathcal{C}_2$. Thus we consider the contracts $\mathcal{C}_2' = (C_2,
\mathbf{p}_2 \setminus \{ x \}, \mathbb{P}'_2)$ and $\mathcal{C}_x = (C_x, \{ x_p \},
\mathbb{P}^x_2)$. $\mathbb{P}'_2$ is the restriction of $\mathbb{P}_2$ without $x$, $\mathbb{P}^x_2$
is the probability distribution considering only $x$, and $C_x$ is a non-probabilistic contract we will call
a {\it wrapper}, with three uncontrolled ports ($x_p$, $x_c$ and $s$) and one controlled port
$x$. This wrapper selects with a non-deterministic port $s \in \{ p,c \}$ the value that will be
given to $x$ between the probabilistic
one and the one given by $\mathcal{C}_1$. Composing $\mathcal{C}'_2$ with $\mathcal{C}_x$ thus
enables us to compose it with $\mathcal{C}_1$ (renaming $x$ to $x_c$). This is illustrated in
Fig.\ref{wrapper}, where thick triangles denote probabilistic ports. The wrong version is on the
top and the correct wrapped one is on the bottom.

We now prove that the composition is compatible with the satisfaction relation. The proof of
this theorem relies on the fact that the probabilistic ports of contracts $\mathcal{C}_1$
and $\mathcal{C}_2$ are disjoint.

\begin{theorem}%
  If $\mathcal{C}_1$ and $\mathcal{C}_2$ are 2 probabilistic contracts that can be composed, $M_1$ and
  $M_2$ 2 implementations such that $ M_i \models_{\beta_i} \mathcal{C}_i \
  \mbox{for} \ i=1,2$ then
  $$M_1 \times M_2 \models_{\beta_1\cdot\beta_2} \mathcal{C}_1 \parallel \mathcal{C}_2.$$
\end{theorem}%

\Notes{Write again this proof differently - done}

\begin{proof}%

The intuition behind this proof is to show that separate histories on the ports $\mathbf{p}_1$ and
$\mathbf{p}_2$, each ensuring that its particular implementation behaves correctly, also ensure that
the composition of the implementations will behave correctly. In short, the product of two correct
histories is correct w.r.t. the composition of the contracts and implementations.
This will be true because every
``correct'' history satisfies the corresponding contract \emph{whatever non-deterministic choices are}.

  Let's consider histories $w_1$ and $w_2$, respectively on the sets $\mathbf{p}_1$ and
  $\mathbf{p}_2$, such that $M_i \cap \{w_i\} \subseteq^{\sigma_i} G_i$. 




As we said before, composition is by intersection over the common ports and variables. We therefore
begin with the inverse projection over the set of variables we want to consider, and then intersect
the runs.

  It is clear that $\forall w, \ \{w\} \cap M_1 \cap M_2 \subseteq^{\sigma} \{w\}
  \cap M_1$, with $\sigma = \sigma_1 \cup \sigma_2 $ (assuming the inverse projections are correctly
  done on both sides). Moreover, $\{ w_1 \times w_2 \} \cap^\sigma M_i \subseteq^\sigma \{w_j\}
  \cap^\sigma M_i$, whatever $i$ or $j$. As a
  consequence:

  $$ \begin{array}{r@{\ }l}
    \{w_1\times w_2\} & \cap M_1 \cap M_2 \subseteq^{\sigma} \{w_1\times w_2\} \cap M_1 \\ 
    & \subseteq^{\sigma} \{w_1\} \cap M_1 \subseteq^\sigma G_1\\
    \\
 
   \{w_1\times w_2\} & \cap M_1 \cap M_2 \subseteq^{\sigma} \{w_1\times w_2\} \cap M_2 \\
   & \subseteq^{\sigma} \{w_2\} \cap M_2 \subseteq^\sigma G_2\\
  \end{array}$$

  $$\Rightarrow \{w_1\times w_2\} \cap M_1 \cap M_2 \subseteq^{\sigma} G_1 \cap G_2$$
  
  As a consequence, $\{w_i\} \cap M_i \subseteq^{\sigma_i} G_i$ implies that 
  $$\{w_1 \times w_2\}
  \cap M_1 \cap M_2 \subseteq^{\sigma_1 \cup \sigma_2} G_1 \cap G_2$$

  And finally

$$ \mathbb{P}(\{w \ | \
    \{w\} \cap M_1 \cap M_2 \subseteq^{\sigma_1 \cup \sigma_2} G_1 \cap G_2 \}) \ge \beta_1 \cdot
    \beta_2 $$

Thus $\mathbb{P}(M_1 \cap M_2 \subseteq \mathcal{C}_1 \parallel \mathcal{C}_2) \ge \beta_1 \cdot
\beta_2$.

\cqfd
\end{proof}

Note that we cannot find a better bound that $\beta_1\cdot\beta_2$,
because if the contracts are independent ($\sigma_1 \cap \sigma_2 =
\emptyset$), we clearly have 
$$\mathcal{M}((M_1\times M_2) \subseteq^{\sigma_1 \cup
    \sigma_2} (G_{C_1} \cap G_{C_2})) = \mathcal{M}(M_1 \subseteq^{\sigma_1} G_{C_1}) \cdot \mathcal{M}(M_2 \subseteq^{\sigma_2} G_{C_2})$$

\subsection{Probabilistic Refinement}

 As we want the probabilistic refinement
 relation to be compatible with composition and satisfaction (and with the non-probabilistic relation), there is not much liberty in the
way we can define it. Let's say that a contract $C_1$ refines a contract $C_2$ (written
$\mathcal{C}_1 \preceq \mathcal{C}_2$). In order to be compatible with the composition, the
probabilistic ports of $\mathcal{C}_1$ must be a subset of those of $\mathcal{C}_2$, and the
distribution on these ports must be the same for $\mathcal{C}_1$ and $\mathcal{C}_2$. Moreover we
want this relation to be compatible with implementation, which means that if an implementation
satisfies $\mathcal{C}_1$ with level $\alpha$, it must satisfy $\mathcal{C}_2$ with a level $\beta$ that can
be computed from $\alpha$. The idea for this is to measure the inclusion of the guarantees of $\mathcal{C}_1$ in the
guarantees of $\mathcal{C}_2$, and to use this measure in order to compute $\beta$.

\begin{definition}
If $\mathcal{C}_1=(C_1,\mathbf{p}_1,\mathbb{P}_1)$ and $\mathcal{C}_2=(C_2,\mathbf{p}_2,\mathbb{P}_2)$ are two
probabilistic contracts, we say that $\mathcal{C}_1$ refines $\mathcal{C}_2$ with level $\gamma$ ($\mathcal{C}_1 \preceq_{\gamma}
\mathcal{C}_2$) if and only if
\begin{enumerate}
\item $\sigma_1 \subseteq \sigma_2$;
\item $\mathbf{p}_1 \subseteq \mathbf{p}_2$ and $\mathbb{P}_1$ is the marginal of $\mathbb{P}_2$ over
  $\mathbf{p}_1$;
\item $\mathbb{P}_2(\{\omega\} \subseteq^{\sigma_2} G_2 | \{\omega\} \subseteq^{\sigma_2} G_1) \geq \gamma$, whatever non-deterministic choices are.
\end{enumerate}
\end{definition}%

$\gamma$ is a measure of the inclusion of $G_1$ in $G_2$, $\gamma = 1$ meaning that $G_1 \subseteq
G_2$ almost all the time.


\Notes{
NB: Perhaps we want to make sure that probabilistic ports and distributions are the same for the
two contracts ?

}

As the definition for the refinement relation was built to be compatible with implementation, it is
quite logical to prove the following theorem, which says that if an implementation satisfies a
probabilistic contract with level $\beta$, and if this contracts refines a second one with level
$\gamma$, then the implementation satisfies the second contract with level $\beta \cdot
\gamma$. This should enable us to use simpler contracts in order to prove satisfaction.

\begin{theorem}%
If $\mathcal{C}_1=(C_1,\mathbf{p}_1,\mathbb{P}_1)$ and $\mathcal{C}_2=(C_2,\mathbf{p}_2,\mathbb{P}_2)$ are 2
probabilistic contracts and $M$ an implementation, then

$$ M \models_\beta \mathcal{C}_1 \ \mbox{and} \ \mathcal{C}_1 \preceq_{\gamma} \mathcal{C}_2 \Rightarrow
M \models_{\beta \cdot \gamma} \mathcal{C}_2.$$
\end{theorem}%

\begin{proof}%
  Because of the definition of the probabilistic refinement, this result is quite clear:
  
  $$\mathbb{P}_2(M \models \mathcal{C}_2) = \mathbb{P}_2(\{w \ | \ \{w\} \cap M \subseteq^{\sigma_2} G_2\})$$
  
  And
  
  $$\begin{array}{c}
    \mathbb{P}_2(\{w \ | \ \{w\} \cap M \subseteq^{\sigma_2} G_2 \})\ge \\
    \mathbb{P}_2(\{w \ | \ \{w\} \cap M
    \subseteq^{\sigma_2} G_1\}) \cdot \mathbb{P}_2( \{w\} \subseteq^{\sigma_2} G_2 \setminus \{w\}
    \subseteq^{\sigma_2} G_1)
  \end{array}$$
  
  And as $\mathbb{P}_1$ is the marginal of $\mathbb{P}_2$ over $\mathbf{p}_1$,
  
  $$\mathbb{P}_2(M \models \mathcal{C}_2) \ge \beta \cdot \gamma$$

\cqfd
\end{proof}%

Once again, we cannot find a finer bound because if $G_2
\subseteq^{\sigma_2} G_1$, we have $\mathbb{P}_2(M \models
\mathcal{C}_2) = \beta\cdot \gamma$.


\subsection{Problems with finer satisfaction relations}

The satisfaction relation we chose above is adapted to reliability analysis. It measures the runs of
the implementations that have the right behaviour \emph{whatever non-deterministic choices
  are}. Consequently one could wonder whether it would be of interest to try finer satisfaction
relations, for example existential or even finer, checking every state the system goes through. 

These finer relations have been studied and ruled out of our work because they cannot be compositional for the following reasons:

\begin{itemize}
\item[$\bullet$] An existential relation would allow us to say that ``in 90\% of the runs, there exist a way in
  which the environment can force our implementation to stay within the bounds of the
  guarantees''. This would be contrary to the principle of re-use, where we want to be sure that
  whatever the user asks of the component, it behaves safely.

\item[$\bullet$] An even finer satisfaction relation that would check every state the system goes
  through would be quite convenient in order to express disponibility properties. This kind of
  satisfaction relation would ponderate each infinite history of the probabilistic ports with the
  ``amount'' of visited states that ensure the guarantees. This would mean, for example, that for a particular
  history on the probabilistic ports, the induced behavior will stay within the guarantees with a
  probability at least $\alpha$, whatever non-deterministic choices are. But knowing this
  probability is not enough to ensure compositionality, because if we compose two contracts, the
  induced behaviour for a fixed history of the probabilistic ports must satisfy both contracts \emph{at the same time}.

\end{itemize}

\subsection{Example}

\begin{figure*}[!tb]
  \begin{center}
    \input{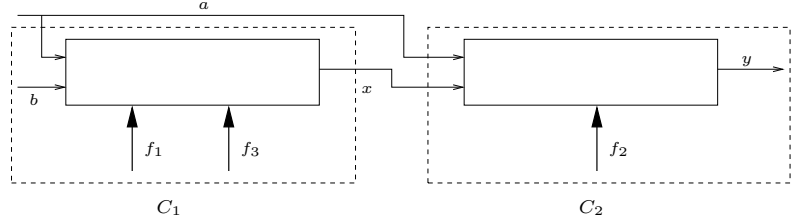}
  \end{center}
  \caption{Global System}
  \label{split_system}
\end{figure*}

Let's assume we want to build a system with $2$ input ports $a$ and $b$, $1$ output port $y$. We want
this system to avoid a state where $y$ is true and $a$ is false. We know there are possibly
different sources of failure $f_1, f_2$ and $f_3$ but we suppose for the moment that $f_3$ will never
happen. We decide to split the system into $2$ subsystems (Fig. \ref{split_system}). We will ask a
first supplier to build the first subsystem as a component satisfying a contract $C_1$, and to a
second supplier, we give a contract $C_2$.




\begin{figure*}
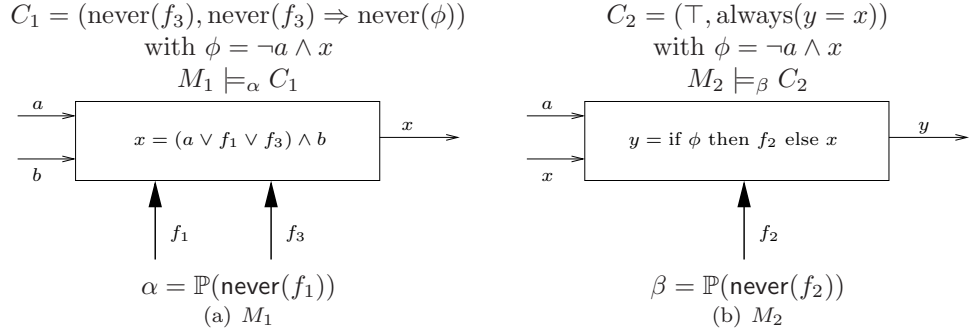

\centering
\mbox{
\subfigure[$M_1$]{\begin{tabular}{c}
    $C_1 = (\mbox{never}(f_3),\mbox{never}(f_3)\Rightarrow \mbox{never}(\phi))$ \\
    with   $\phi = \lnot a \land x$    \\
    $M_1 \models_\alpha C_1$ \\
    \input{C1.pstex_t} \\
    $\alpha = \mathbb{P}(\mathsf{never}(f_1))$
  \end{tabular}\label{m1}}\quad
\subfigure[$M_2$]{\begin{tabular}{c}
    $C_2 = (\top,\mbox{always}(y=x))$ \\
    with   $\phi = \lnot a \land x$    \\
    $M_2 \models_\beta C_2$ \\
     \input{C2.pstex_t} \\
     $\beta = \mathbb{P}(\mathsf{never}(f_2))$
  \end{tabular}\label{m2}}
}
\caption{Subcomponents}
\label{contracts12}
\end{figure*}

\begin{figure*}
\centering
\begin{tabular}{c}
 $C = C_1 \parallel C_2 =
(\mbox{never}(f_3),\mbox{never}(f_3) \Rightarrow \mbox{never}(\lnot a \land y))$
\\
$M_1 \parallel M_2 \models_{\alpha \cdot \beta} C$
\\
  \input{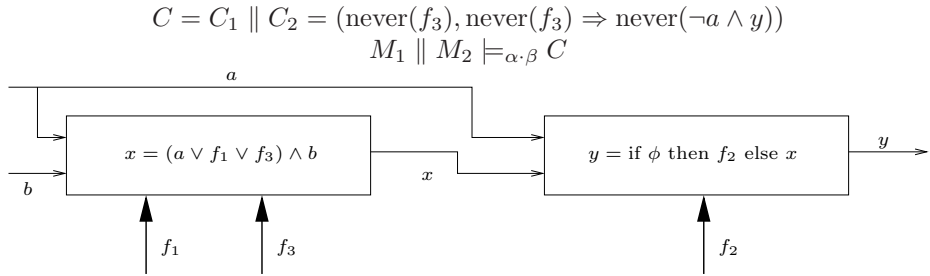}

\end{tabular}

  \caption{Composition of the implementations}
  \label{composition}
\end{figure*}

The first supplier will then provide us with an implementation $M_1$ satisfying $C_1$ with a level
$\alpha$ (Fig. \ref{m1}), and
the second will give us an implementation $M_2$ satisfying $C_2$ with a level $\beta$ (Fig. \ref{m2}). Once we have these
components, we know that their composition will satisfy the contract $C = C_1 \parallel C_2 =
(\mbox{never}(f_3),\mbox{never}(f_3) \Rightarrow \mbox{never}(\lnot a \land y))$, with a level $\alpha \cdot \beta$~(Fig.~\ref{composition}).

Now consider the case when we discover that $f_3$ may in fact happen. The contract $C$ is not
realistic anymore, as it supposes that $f_3$ never happens. In consequence, we want to know how our
components will satisfy a contract $C' = (\top, \mbox{never}(\lnot a \land y))$. Instead of trying to
find a new decomposition into different subcontracts, we just have to compute the level of
refinement $\gamma$ such that $ C \preceq_\gamma C'$. We will then know that $M_1 \parallel M_2
\models_{\alpha\beta\gamma} C'$. This probability $\gamma$ may be written as follows:

  $$\gamma = \mathbb{P}(\mbox{never}(\lnot a \land y) | \mbox{never}(f_3) \Rightarrow \mbox{never}(\lnot a \land y))$$

\section{Related work}

The problem of reliability analysis is widely present in the literature. Several attempts have been
presented in the domain of probabilistic model checking in order to express probabilistic properties
and check whether a particular system satisfies them. pCTL and pCTL$^*$, for example, can be used to
specify properties such as reliability and performance~\cite{HJ89,ASSB95}. There even exist extensions of pCTL and
pCTL$^*$ in which the probabilistic behaviour coexist with non-determinism~\cite{BA95}.
However, in these formalisms, the probabilistic point of view is inherent to the system
checked. Consequently, compositionality is not an issue for them. 
Our formalism, on the contrary, considers probabilities as an assumption on the environment. In this
way, we only consider open systems for which probabilities and non-determinism comes
from the environment. In this way, compositionality can be proved and used in order to obtain a
modular analysis.

On the other hand, compositional reliability analysis tools and formalisms have already been
developed in the literature, such as
Arcade~\cite{BoudaliICECCS08} or Dynamic Fault
Trees~\cite{BoudaliCS07} for example. These formalisms present compositional reliability analysis as
it is actually done in the industry, that is to say without any behavioural interpretation. Our
approach is different. We want to be able to reason on the behaviours of components, and not only on
their failure probability. Of course our formalism captures such classical analysis, but it allows
much more because the satisfaction relation is strongly linked to the behaviour of the
implementations. Moreover, as our formalism allows assumptions on the environment, it can capture
situations where two separate implementations do not satisfy their respective contracts, but their
composition satisfies the composition of the contracts because of the assumptions on the
environment, which would not be possible with a classical reliability analysis.

Finally, the probabilistic refinement relation we have built does not have an equivalent in the
classical reliability analysis. It allows to compute the probabilistic satisfaction of a contract
while only considering information on the probabilistic satisfaction of another contract and on the
relations between these contracts.

\section{Conclusion and further work}

In this paper, we have presented a compositional theory of probabilistic contracts, capable of
capturing reliability aspects of components and systems. This theory enables a behavioural
interpretation of reliability, which was not the case in the existing compositional formalisms.

There are several natural directions to continue this work. 

First, what we present here is a very general theory with few direct applications. Computing the
satisfaction and refinement probabilities efficiently would require to narrow the field of
applications.
In practice, assertions and machines will be deterministic open transition systems and never sets of run. We are
actually developing a more practical approach where contracts are Markov Decision Processes and
implementations open transition systems. In this approach, computing the satisfaction and refinement
probabilities relies on the existence of pure optimal strategies in mean-payoff Markov Decision Processes~\cite{Gimbert07}.

Finally, the same kind of probabilistic point of view could be adapted
to contracts residuation~\cite{Raclet08},
which would give a practical way to build (canonical?) implementations from the residuation of the
guarantees of a contract by its assumptions.

\bibliographystyle{amsalpha}
\bibliography{bibliography}

\tableofcontents

\end{document}